\documentclass[12pt]{iopart}
\usepackage{psfig,epsfig}

\def\gapprox{\lower.4ex\hbox{$\;\buildrel >\over{\scriptstyle\sim}\;$}}
\def\lapprox{\lower.4ex\hbox{$\;\buildrel <\over{\scriptstyle\sim}\;$}}
\def\half{{\textstyle{1\over2}}}

\def\be{\begin{equation}}
\def\ee{\end{equation}}
\def\bea{\begin{eqnarray}}
\def\nn{\nonumber}
\def\eea{\end{eqnarray}}
\newcommand{\msp}{\noalign{\vspace{3pt plus2pt minus1pt}}}

\def\bi{\bf}
\def\p{{\rm p}}
\def\rmF{{\rm F}}
\def\ve{\varepsilon}

\def\pf{p_\rmF}
\def\vef{\ve_\rmF}
\def\bkm{|{\bf k}|}
\def\etal{{\it et al} }

\begin{document}

\title[Relativistic quantum plasma dispersion functions]
{Relativistic quantum plasma dispersion functions}

\author[D B Melrose, J I Weise and J McOrist]%
{D B Melrose$^1$, J I Weise$^1$ and J McOrist$^{1,2}$}
\address{
$^1$School of Physics, University of Sydney, NSW 2006, AUSTRALIA\\
$^2$Department of Physics, University of Chicago, 5640 S. Ellis Ave.,
Chicago, IL 60637, USA}
\ead{melrose@physics.usyd.edu.au}

\begin{abstract}
Relativistic quantum plasma dispersion functions are defined and the
longitudinal and transverse response functions for an electron (plus
positron) gas are written in terms of them. The dispersion is
separated into Landau-damping, pair-creation and dissipationless
regimes. Explicit forms are given for the RQPDFs in the cases of a
completely degenerate distribution and a nondegenerate thermal
(J\"uttner) distribution. Particular emphasis is placed on the
relation between dissipation and dispersion, with the dissipation
treated in terms of the imaginary parts of RQPDFs. Comparing the
dissipation calculated in this way with the existing treatments leads
to the identification of errors in the literature, which we correct.
We also comment on a controversy as to whether the dispersion curves
in a superdense plasma pass through the region where pair creation is
allowed.
\end{abstract}


\maketitle

\section{Introduction}

There is a relatively sparse but diverse body of literature on
relativistic quantum plasma dispersion functions (RQPDFs). Tsytovich
(1961) calculated the response of a relativistic quantum (RQ)
electron gas and derived expressions for the real and imaginary
parts of the longitudinal and transverse response functions for
arbitrary isotropic distributions, and for a J\"uttner distribution.
Jancovici (1962) derived expressions for the real and imaginary
parts of the longitudinal and transverse response functions for a
completely degenerate electrons gas, providing a relativistic
generalization of the well-known result of Lindhard (1954) for a
nonrelativistic degenerate electron gas. These results were
rederived and extended in various ways, by Hakim and Heyvaerts
(1978, 1980) and Sivak (1985), using a Wigner function approach
(Hakim 1978), by Delsante and Frankel (1980) and Kowalenko {\it et
al}  (1985), who concentrated on the longitudinal response, and by
Hayes and Melrose (1984), who derived general results and Melrose
and Hayes (1984),  who extended Jancovici's (1962) results to the
nearly degenerate limit and Tsytovich's (1961) results to a mildly
degenerate plasma. The response functions were discussed further by
Itoh \etal (1992) and Braaten and Segel (1993) in connection with
neutrino losses from stellar interiors. This area is of ongoing
interest  (Ratkovi$\acute c$, Dutta and Prakash 2003; Dutta,
Ratkovi$\acute c$ and Prakash 2004; Koers and Wijers 2005; Jaikumar,
Gale and Page 2005). One of the neutrino emission processes, the
so-called plasma process, is dependent on dispersion and dissipation
in the plasma, and the results of Braaten and Segel (1993) are used.
Although these authors included RQ effects in their formal
development, they approximated the general result, effectively by
neglecting RQ effects in the resonant denominator, and it is their
approximate form that is used. We argue that this leads to a
misleading conclusion concerning one-photon pair creation (PC) in a
superdense plasma, which we define here as plasmas with cutoff
frequency (called the plasma frequency by some authors) exceeding
the threshold, $2m$, for PC. (We use natural units with
$\hbar=c=1$.)

Our main purpose in this paper is to discuss the properties of RQPDFs
in an isotropic, thermal RQ electron gas emphasizing the relation
between dispersion and dissipation and the role of PC. One general
feature of existing treatments is that the dispersion and dissipation
are treated in different ways, with the dispersion described by
appropriate RQPDFs, and with the dissipation calculated directly from
the resonant part of the response function. In principle, the RQPDFs
have imaginary parts that describe the dissipation, with the relation
between the real and imaginary parts determined by the causal
condition (e.g., the Landau criterion). Calculation of the
dissipative part in this way provides a useful consistency check on
the expression for the real part of the response function. Our
consistency check fails in two published cases, and we identify and
correct the relevant errors.

There are relatively few applications where the combination of
intrinsically relativistic and quantum effects is important in a
plasma. The extreme conditions required apply in, for example, the
early Universe, quark-gluon plasma and the interiors of compact
stars.  In section~6 we discuss possible applications, emphasizing a
specific point relevant to the plasma process for neutrino emission:
whether or not one-photon PC is possible is a superdense plasma.
Earlier authors (Tsytovich 1961, Hayes and Melrose 1984, Kowalenko
{\it et al}  1985) assumed that the dispersion curve does pass
through the PC region, so that PC is allowed, and Itoh \etal (1992)
and Braaten and Segel (1993) gave arguments against this. Here we
show that PC is possible in a superdense plasma and we determine the
conditions under which it can occur.

In section~2 we present general formulae for the response functions
and define general forms of RQPDFs for isotropic distributions. In
section~3 we discuss dissipation, giving particular emphasis to the
boundaries of the regions when Landau damping (LD) and PC are
allowed for a given particle. In sections~4 and~5 we discuss the
responses of a completely degenerate distribution and a
nondegenerate thermal distribution, respectively. We discuss
applications in section~6 and summarize our conclusions in
section~7.

\section{RQPDFs for isotropic distributions}

An isotropic distribution is isotropic in one inertial frame, which
is the rest frame of the medium. The linear response tensor for an
isotropic medium may be described in terms of the longitudinal and
transverse response functions.  In this section, we start with a
general form for the response tensor, then write down explicit forms
for the longitudinal and transverse response functions, and identify
relevant RQPDFs.

\subsection{General form for the response tensor}

The general expression for the response tensor has been written down
in a variety of different forms. We start with a covariant form that
is derived by analogy with the (unregularized) vacuum polarization
tensor (e.g., Berestetskii {\it et al} 1971), ${\cal P}^{\mu\nu}(k)$,
which relates the Fourier transform in space and time of the linear
induced 4-current, $J^\mu(k)$, to the 4-potential, $A^\mu(k)$, where
$k$ denotes the wave 4-vector, $[\omega,{\bf k}]$, constructed from
the frequency, $\omega$, and the wave 3-vector, $\bf k$. The response
4-tensor satisfies the charge-continuity and gauge-invariance
relations, $k_\mu{\cal P}^{\mu\nu}(k)=0$, $k_\nu{\cal
P}^{\mu\nu}(k)=0$. This form is
\bea
\fl
{\cal P}^{\mu\nu}(k)=
-e^2\sum_{\zeta,\zeta'}
\int{\rmd^3{\bi p}\over(2\pi)^3}
\int{\rmd^3{\bi p}'\over(2\pi)^3}\,
(2\pi)^3\delta^3(\zeta'{\bi p}'-\zeta{\bi p}+{\bi k})\,
\nonumber
\\
\qquad\qquad\qquad
\times
{\half(\zeta'-\zeta)+\zeta n^\zeta({\bi p}) -\zeta'n^{\zeta'}({\bi p}')
\over
\omega-\zeta\varepsilon+\zeta'\varepsilon'}\,
{F^{\mu\nu}(\zeta{\tilde p},
\zeta'{\tilde p}')
\over\zeta\varepsilon\,\zeta'\varepsilon'},
\qquad\qquad
\label{calP0}
\eea
where $n^\zeta({\bi p})$ is the occupation number, with
$\zeta,\zeta'=\pm1$ labeling electron and positron states, and with
\be
F^{\mu\nu}(P,P')
=P^\mu P'^\nu+P'^\mu P^\nu
+g^{\mu\nu}(m^2-PP').
\label{Fmunu}
\ee
The energies are $\ve=\ve({\bi p})=(m^2+|{\bi p}|^2)^{1/2}$,
$\ve'=\ve({\bi p}')$, and ${\tilde p}$ denotes the 4-momentum with
components $[\ve,{\bi p}]$. The (unregularized) vacuum polarization
tensor itself follows from (\ref{calP0}) by neglecting the
contribution of the particles ($n^\zeta({\bi p})\to0$).

\subsection{Longitudinal and transverse response functions}

Isotropy implies that the response tensor is of the form
\be
{\cal P}^{\mu\nu}(k)={\cal P}^L(k)\, L^{\mu\nu}(k,{\tilde u})+
{\cal P}^T(k) \,T^{\mu\nu}(k,{\tilde u}),
\label{LTsep}
\ee
where $L^{\mu\nu}(k,{\tilde u})$  and $T^{\mu\nu}(k,{\tilde u})$ are
longitudinal and transverse  projection operators that depend on the
4-velocity ${\tilde u}^\mu$ of the rest frame of the plasma. After
projecting (\ref{calP0}) to identify ${\cal P}^L(k)$ and ${\cal
P}^T(k)$, one may perform the ${\bi p}'$ integral over the
$\delta$-function, choose the rest frame and rewrite the ${\bi
p}$-integral in terms of integrals over $\ve$, $\ve'$:
$$\int\rmd^3{\bi p}\to2\pi\int_0^\infty\rmd|{\bi p}|\,|{\bi
p}|^2\int_{-1}^1\rmd\cos\theta
={2\pi\over|{\bi k}|}\int_m^\infty\rmd\varepsilon\,\varepsilon
\int_{\varepsilon'_{\rm min}}^{\varepsilon'_{\rm
max}}\rmd\varepsilon'\,\varepsilon',
$$
where the limiting values are
\be
\varepsilon'_{\rm max,min}=(\varepsilon^2
\pm2|{\bi p}||{\bi k}| +|{\bi k}|^2)^{1/2}.
\label{limits}
\ee
The response functions are unchanged by interchanging electrons and
positrons. One may evaluate the response functions for electrons, and
then replace the occupation number by
\be
{\tilde n}(\ve)=n^+(\ve)+n^-(\ve),
\label{tilden}
\ee
to include the contribution of the positrons.

Explicit forms for the response functions (neglecting the vacuum
contribution) are (Hayes and Melrose 1984)
\be
\fl
{\cal P}^L(k)=
{e^2n_{\p 0}\omega^2\over m|{\bi k}|^2}
+{e^2\omega^2m\over8\pi^2|{\bi k}|^3}\,
\bigg[
(\omega^2-|{\bi k}|^2)
S^{(0)}(k)
-4m\omega S^{(1)}(k)
+4m^2S^{(2)}(k)
\bigg],
\label{calPL}
\ee
\bea
\fl
{\cal P}^T(k)=
-{e^2n_{\p 0}(\omega^2+|{\bi k}|^2)\over2m|{\bi k}|^2}
-{e^2(\omega^2-|{\bi k}|^2)m\over16\pi^2|{\bi k}|^3}
\bigg[
(-4\varepsilon_k^2
+\omega^2+2|{\bi k}|^2)S^{(0)}(k)
\nonumber
\\
\qquad\qquad
\qquad\qquad\qquad\qquad
-4m\omega S^{(1)}(k)
+4m^2S^{(2)}(k)
\bigg],
\label{calPT}
\eea
with $n_{\p 0}$ the proper number density, with
\be
\ve_k={|{\bf k}|\over2}\left({\omega^2-|{\bf
k}|^2-4m^2\over\omega^2-|{\bf k}|^2}
\right)^{1/2},
\label{vek}
\ee
and where three RQPDFs are introduced, $S^{(n)}(k)$, with $n=0,1,2$.
These RQPDFs involve integrals over the occupation number and two
logarithmic functions:
\be
\fl
S^{(n)}(k)=
\int{\rmd\varepsilon\over m}\,\left({\varepsilon\over m}\right)^{n}
{\tilde n}(\varepsilon)
\cases{\ln\Lambda_1
& for $n=0,2$,
\cr
\ln\Lambda_2
&for $n=1$,
\cr}
\label{Sn1}
\ee
\be
\fl
\Lambda_1=
{4\varepsilon^2\omega^2-
(\omega^2-|{\bi k}|^2-2|{\bi p}|\,|{\bi k}|)^2
\over
4\varepsilon^2\omega^2-
(\omega^2-|{\bi k}|^2+2|{\bi p}|\,|{\bi k}|)^2},
\quad
\Lambda_2=
{4(\varepsilon\omega+|{\bi p}|\,|{\bi k}|)^2
-(\omega^2-|{\bi k}|^2)^2
\over
4(\varepsilon\omega-|{\bi p}|\,|{\bi k}|)^2
-(\omega^2-|{\bi k}|^2)^2}.
\label{Lambdas}
\ee

The response functions ${\cal P}^{L,T}$ are related to the dielectric
response functions $\epsilon^{L,T}$ of Jancovici (1962) and Kowalenko
{\it et al} (1985) as follows:
\begin{equation}
{\rm Im}\, \epsilon^{L,T}={\mu_0\over\omega^2}{\rm Im}\,  {\cal P}^{L,T},\qquad
{\rm Re}\,\epsilon^{L,T}=1+{\mu_0\over\omega^2}{\rm Re}\,{\cal P}^{L,T},
\end{equation}
resulting in dispersion relations in the rest frame of the plasma of the form
\begin{equation}
\omega^2+\mu_0\,{\rm Re}\,{\cal P}^L(k)=0,
\qquad
\omega^2-|{\bf k}|^2+\mu_0\,{\rm Re}\,{\cal P}^T(k)=0.
\end{equation}

\section{Dissipation}

As already noted, in the existing literature it has been conventional
to treat dissipation separately from dispersion, rather than treating
it in terms of the imaginary parts of RQPDFs. Dissipation is possible
only when $\ve_k$, as given by (\ref{vek}), is real:  LD applies for
$\omega^2-|{\bf k}|^2<0$ and PC for $\omega^2-|{\bf k}|^2>4m^2$.
Before discussing the conventional procedure, it is useful to
identify the energy, momentum and speed of a resonant particle at the
boundary of the allowed regions for LD or PC.

\subsection{Limiting values of the resonance condition}

In dispersion theory it is conventional to refer to a zero of the
denominator (as a function of $\omega$ for fixed ${\bi k}$) as a
resonance, and to the algebraic condition for such a zero as a
resonance condition. The Landau prescription specifies how one is to
integrate around the associated pole in the integrand in accord with
the causal condition, such that each pole contributes an imaginary
part equal to its semi-residue. Dissipation is described by the
contributions from these semi-residues, and different poles are
interpreted in terms of different dissipation processes. There are
two dissipation processes in a collisionless, unmagnetized plasma, LD
and PC.

For given $\omega,|{\bi k}|$, the limiting values of the resonance
condition determine discrete values of $\ve,|{\bi p}|,|{\bi v}|$.
These values are found by setting
$\omega-\zeta\varepsilon+\zeta'\varepsilon'=0$, squaring twice to
remove the square roots, and setting $({\bi p}\cdot{\bi k})^2=|{\bi
p}|^2|{\bi k}|^2$. Writing $|{\bi p}|=m\sinh\chi$,
$\varepsilon=m\cosh\chi$, $|{\bi v}|=\tanh\chi$,
$t=\tanh(\half\chi)$, the limiting values must satsify
\be
\fl
(1+a)^2t^4+2(1-a^2-2b^2)t^2+(1-a)^2=0,
\qquad
a={\omega^2-|{\bi k}|^2\over2m\omega},
\quad
b={|{\bi k}|\over\omega}.
\label{ab}
\ee
The solutions for $t^2$ are
\be
t^2=t_\pm^2,
\qquad
t_\pm={b\pm(a^2+b^2-1)^{1/2}
\over1+a}.
\label{tpm}
\ee
There is considerable freedom in choosing the four solutions. An
obvious choice is $t=\pm t_+$, $t=\pm t_-$. Noting that $t$ and
$-1/t$ correspond to the same values of $\ve,|{\bi p}|,|{\bi v}|$,
one may also choose from $t=\pm1/t_+$, $t=\pm1/t_-$. With $t=t_\pm$,
the solutions for the energy, momentum and speed are
\bea
{\varepsilon_\pm\over m}={a\pm b(a^2+b^2-1)^{1/2}
\over1-b^2},
\qquad
{p_\pm\over m}={ab\pm (a^2+b^2-1)^{1/2}
\over1-b^2},
\nn
\\
\msp
v_\pm={p_\pm\over\varepsilon_\pm}
={b\pm a(a^2+b^2-1)^{1/2}
\over a^2+b^2}.
\qquad\qquad\qquad\qquad
\label{n4}
\eea
Another form of the boundary solutions was introduced by Tsytovich
(1961); these are related to the $\pm$-solutions by
\be
\ve_\pm=\half\omega\pm\ve_k,
\qquad
p_\pm=\half|{\bi k}|\pm\frac{\omega}{|{\bi k}|}\,\ve_k,
\label{vekpm}
\ee
with $\ve_k$ given by (\ref{vek}). The solutions (\ref{vekpm}) are
the natural solutions for PC, and they correspond to the choice
$t=t_\pm$. For LD, the natural solutions are
$\ve=\ve_k\pm\half\omega$, $|{\bi p}|=\omega\ve_k/\bkm\pm\half\bkm$,
and these correspond to $t=t_-$ and $t=1/t_+$, respectively. At a
boundary of the LD or PC region, $\ve$ must correspond to either
$\ve_\pm$, with $\ve'_{\rm max,min}$ then corresponding to $\ve_\mp$,
but some care is required in making the specific identifications.

\subsection{Dissipation due to LD and PC}

The resonant terms in (\ref{calP0}) correspond to replacing the
denominator by $-\rmi\pi \delta(\omega-\zeta\ve+\zeta'\ve')$, where
the Landau prescription is used. We discuss LD and PC separately.

The terms with $\zeta=\zeta'=\pm1$ describe LD, and these correspond
to resonances at $\omega=\pm(\ve-\ve')$, respectively. On repeating
the derivation of the longitudinal and transverse parts, the double
integral over $\ve$, $\ve'$ is reduced to a single integral by this
$\delta$-function, and the limit of integration can be expressed in
terms of $\ve_k$. In this way Tsytovich (1961) derived imaginary
parts that correspond to
\be
\fl
{\rm Im}\,{\cal P}^L_{\rm LD}(k)=
{e^2\omega^2\over8\pi|{\bi k}|^3}
\int_{\varepsilon_k}^{\infty}\rmd\varepsilon''\,
(4\ve''^2-|{\bi k}|^2)
[{\tilde n}(\varepsilon''-\half\omega)-{\tilde n}(\varepsilon''+\half\omega)],
\label{ImPLLD}
\ee
\be
\fl
{\rm Im}\,{\cal P}^T_{\rm LD}(k)=
{e^2(|{\bi k}|^2-\omega^2)\over8\pi|{\bi k}|^3}
\int_{\varepsilon_k}^{\infty}\rmd\varepsilon''\,
(2\ve''^2-2\ve_k^2+|{\bi k}|^2)
[{\tilde n}(\varepsilon''-\half\omega)-{\tilde n}(\varepsilon''+\half\omega)],
\label{ImPTLD}
\ee
in the notation used in this paper.

The resonant part of (\ref{calP0}) describes PC for $\zeta\zeta'=-1$,
but only $\zeta=1=-\zeta'$ contributes for $\omega>0$. The vacuum
contributes in this case, and Tsytovich (1961) retained both the
contribution of the vacuum and of the electron gas, but his final
expression for the vacuum contribution differs from the well-known
result by a factor of two. We note that the response is invariant
under the interchange of positrons and electrons, which allows one to
rewrite $1-n^+(\ve)-n^-(\ve')$ in (\ref{calP0}) for $\zeta=1=-\zeta'$
as $1-\half{\tilde n}(\ve)-\half{\tilde n}(\ve')$. Then, repeating
the derivation, we find
\be
\fl
{\rm Im}\,{\cal P}^L_{\rm PC}(k)=
-{e^2\omega^2\over8\pi|{\bi k}|^3}
\int_{-\varepsilon_k}^{\varepsilon_k}\rmd\varepsilon''\,
(4\ve''^2-|{\bi k}|^2)
[1-\half{\tilde n}(\half\omega+\varepsilon'')-\half{\tilde
n}(\half\omega-\varepsilon'')],
\label{ImPLPC}
\ee
\be
\fl
{\rm Im}\,{\cal P}^T_{\rm PC}(k)=
{e^2(\omega^2-|{\bi k}|^2)\over8\pi|{\bi k}|^3}
\int_{-\varepsilon_k}^{\varepsilon_k}\rmd\varepsilon''\,
(2\ve''^2-2\ve_k^2+|{\bi k}|^2)
[1-\half{\tilde n}(\half\omega+\varepsilon'')-\half{\tilde
n}(\half\omega-\varepsilon'')],
\label{ImPTPC}
\ee
for the imaginary parts due to PC of the longitudinal and transverse
responses, respectively. The unit term inside the square brackets
differs by the relevant factor of two from Tsytovich's expression.

\subsection{Dissipation due to the vacuum polarization tensor}

The vacuum polarization tensor is of the form ${\cal
P}_0^{\mu\nu}(k)={\cal P}_0(k)(g^{\mu\nu}-k^\mu k^\nu/k^2)$, where
${\cal P}_0(k)$ is an invariant. The longitudinal and transverse
parts are ${\cal P}_0^L(k)=\omega^2{\cal P}_0(k)/(\omega^2-|{\bf
k}|^2)$ and ${\cal P}_0^T(k)={\cal P}_0(k)$, respectively. The real
part is negligible for most purposes involving wave dispersion, but
the imaginary part cannot be neglected when considering dissipation
due to PC. An important point is that dissipation due to PC occurs in
the vacuum, and the presence of an electron gas tends to suppress it
due to the Pauli exclusion principle. The imaginary part of the
vacuum polarization tensor is well-known, e.g., Berestetskii {\it et
al}  (1971), and corresponds to
\be
{\rm Im}\,{\cal P}_0(k)=
{e^2\over3\pi}\,
\big[m^2+\half(\omega^2-|{\bi k}|^2)\big]\,\frac{\ve_k}{\bkm},
\label{Imvac}
\ee
for $\omega^2-|{\bi k}|^2>4m^2$, with ${\rm Im}\,{\cal P}_0(k)=0$ for
$\omega^2-|{\bi k}|^2<4m^2$. A derivation of (\ref{Imvac}) using the
approach adopted here leads to (\ref{ImPLPC}) and (\ref{ImPTPC}) with
only the unit terms retained. The integrals are then elementary and
(\ref{ImPLPC}) and (\ref{ImPTPC}) reproduces the longitudinal and
transverse parts of (\ref{Imvac}), respectivley.

\subsection{Imaginary parts of logarithmic PDFs}

Logarithmic functions appear naturally in (\ref{Sn1}), and more
specifically as RQPDFs for a completely degenerate electron gas, as
discussed below. It is desirable to have a prescription that allows
one to write down the imaginary part of a logarithmic PDF directly.
Superficially, this seems trivial: when $x$ passes from positive to
negative, $\ln x$ may be replaced by $\ln|x|\pm \rmi\pi$. However,
determining the relevant sign is not trivial.

The imaginary part of any PDF may be determined by imposing the
Landau prescription. For a logarithmic PDF, this leads to the generic
prescription
\be
\fl
{\rm PDF}(\omega)=-\int_{\omega_{\rm min}}^{\omega_{\rm max}}
\frac{\rmd x}{\omega-x+\rmi0}=
\left\{
\!\!\!
\begin{array}{ll}
\ln
\displaystyle{\omega-\omega_{\rm max}\over\omega-\omega_{\rm min}}
&{\rm for}\ \omega<\omega_{\rm min}, \quad \omega>\omega_{\rm max},
\\
\msp
\ln\left|
{\displaystyle
{\omega-\omega_{\rm max}\over\omega-\omega_{\rm min}}
}\right|
+\rmi\pi
&{\rm for}\ \omega_{\rm min}< \omega<\omega_{\rm max}.
\end{array}
\right.
\label{Imln}
\ee
Hence, to impose the causal condition, one writes a logarithmic
function as a sum (or difference) of terms of the form (\ref{Imln})
and gives each term an imaginary part, $\rmi\pi$, when the frequency
is in the range $\omega_{\rm min}< \omega<\omega_{\rm max}$.

\section{Completely degenerate Fermi gas}

The response of a completely degenerate Fermi gas was calculated by
Jancovici (1962), cf.\ also Hayes and Melrose (1984), Sivak (1985),
Kowalenko {\it et al}  (1985).  Jancovici's expression for the
transverse part of the response tensor contains a spurious factor
$\omega^2/(\omega^2-|{\bf k}|^2)$, which leads to a nonphysical
resonance in the dispersion relation for transverse waves. It also
implies incorrectly that the contribution of the electron gas to
dissipation due to LD and PC has the same sign for the transverse
response. Here we start with forms that are valid in the DL region,
where the response functions are necessarily real, and then discuss
the extensions into the LD and PC regions.

\subsection{Thermal distributions}

Before considering the completely degenerate limit, it is appropriate
to comment on the general case of a thermal distribution of
electrons, which is the Fermi-Dirac distribution,
\be
{\tilde n}(\varepsilon)={1\over \exp[(\varepsilon-\mu)/T]+1}
+{1\over \exp[(\varepsilon+\mu)/T]+1},
\label{FD}
\ee
where the chemical potential, $\mu$, includes the rest energy, $m$,
and where the temperature is $T$, with Boltzmann's constant set to
unity. (The form (\ref{FD}) applies for each spin state, and a factor
of two arises when one sums over the two spin states for unpolarized
electrons and positrons.) No general results are known for the RQPDFs
for this case. The completely degenerate limit corresponds to
$T\to0$, $\mu\to\ve_\rmF$, when one has
\be
{\tilde n}(\varepsilon)=\cases{1& for $\ve<\ve_\rmF$,
\cr
\msp
0 & for $\ve>\ve_\rmF$,
\cr}
\label{degenerate}
\ee
where $\ve_\rmF=(m^2+p_\rmF^2)^{1/2}$ is the Fermi energy, with the
Fermi momentum determined by the electron number density,
$n_e=p_\rmF^3/3\pi^2$.

\subsection{DL region}

In the DL region Jancovici's response functions, in the present
notation (with the spurious factor omitted), are
\bea
\fl
{\cal P}^L(k)={e^2\omega^2\over4\pi^2|{\bi k}|^2}\,
\bigg\{{8\varepsilon_{\rm F}p_{\rm F}\over3}
-{2|{\bi k}|^2\over3}\ln\left(
{\varepsilon_{\rm F}+p_{\rm F}
\over m}
\right)
+{\varepsilon_{\rm F}
[4\varepsilon_{\rm F}^2+3(\omega^2-|{\bi k}|^2)]\,
\over6|{\bi k}|}\,
\ln\Lambda_{1\rmF}
\nonumber
\\
+{\omega[3|{\bi k}|^2-\omega^2-12\varepsilon_{\rm F}^2]
\over12|{\bi k}|}
\ln\Lambda_{2\rmF}
+{2m^2+\omega^2-|{\bi k}|^2\over3(\omega^2-|{\bi k}|^2)}
|{\bi k}|\varepsilon_k\,
{\omega\over|\omega|}\,
\ln\Lambda_{3\rmF}
\bigg\},
\label{JL}
\eea
\bea
\fl
{\cal P}^T(k)=-{e^2\over4\pi^2}\,
\bigg\{{4\omega^2+2|{\bi k}|^2\over3|{\bi k}|^2}\,
\varepsilon_{\rm F}p_{\rm F}
+{2(\omega^2-|{\bi k}|^2)\over3}\ln\left(
{\varepsilon_{\rm F}+p_{\rm F}
\over m}
\right)
\nonumber
\\
\qquad
+\varepsilon_{\rm F}
\left[{\varepsilon_{\rm F}^2(\omega^2-|{\bi k}|^2)\over3|{\bi k}|^3}
+{4m^2|{\bi k}|^2+\omega^4-|{\bi k}|^4\over4|{\bi k}|^3}
\right]\,
\ln\Lambda_{1\rmF}
\nonumber
\\
\qquad\qquad
-{\omega
[12m^2|{\bi k}|^2
+(\omega^2-|{\bi k}|^2)(12\varepsilon_{\rm F}^2
+\omega^2+3|{\bi k}|^2)]
\over24|{\bi k}|^3}\,
\ln\Lambda_{2\rmF}
\nonumber
\\
\qquad\qquad\qquad
-{2m^2+\omega^2-|{\bi k}|^2\over3|{\bi k}|}\,
\varepsilon_k\,
{\omega\over|\omega|}\,
\ln\Lambda_{3\rmF}
\bigg\},
\label{JT}
\eea
with $\Lambda_{i\rmF}$ given by setting $|{\bf p}|=p_\rmF$,
$\ve=\ve_\rmF$ in the expressions (\ref{Lambdas}) for  $\Lambda_1$,
$\Lambda_2$ and
\be
\Lambda_3={(\omega^2-|{\bi k}|^2)^2
(\varepsilon|{\bi k}|+2|{\bi p}|\varepsilon_k)^2
-4m^4\omega^2|{\bi k}|^2
\over
(\omega^2-|{\bi k}|^2)^2
(\varepsilon|{\bi k}|-2|{\bi p}|\varepsilon_k)^2
-4m^4\omega^2|{\bi k}|^2}.
\label{Lambda3}
\ee
Note that the sign $\omega/|\omega|$ is needed in the terms involving
$\Lambda_{3\rmF}$ to ensure that the real parts of the response
functions are even functions of $\omega$.

The DL regime corresponds to $|{\bf k}|^2<\omega^2<4m^2+|{\bf k}|^2$.
In this regime, $\Lambda_{1\rmF}$ and $\Lambda_{2\rmF}$ are positive
(they are always real) so their logarithms are real. Due to
$\varepsilon_k$ being imaginary, $\Lambda_{3\rmF}$ is the ratio of a
complex number and its complex conjugate, and hence its logarithm is
imaginary, so that $\varepsilon_k\ln\Lambda_{3\rmF}$ is real. One may
write
\be
\varepsilon_k\ln\Lambda_{3\rmF}=-2|\varepsilon_k|\cases{\arctan\chi_+-\arctan\chi_-
& for $ \chi_->0$,\cr
\arctan\chi_+-\arctan\chi_--\pi
&for $\chi_-<0$,\cr}
\label{Lambda3F}
\ee
\be
\chi_\pm=\frac{2\vef(\omega^2-|{\bf k}|^2)|\ve_k|}
{|{\bf k}|[2m^2|{\bf k}|\pm \pf(\omega^2-|{\bf k}|^2)]}.
\label{chipm}
\ee
Although earlier authors, e.g., Kowalenko {\it et al}  (1985), Sivak
(1985), noted that $\ln\Lambda_{3\rmF}$ is replaced by an arctangent,
there are many choices as to how this arctangent is written. With the
choice made in (\ref{Lambda3F}) the arctangents $\chi_\pm$ remain
between 0 and $\pi/2$ throughout the DL range. The choice
(\ref{chipm}) avoids complications with other choices in numerical
calculations.

\subsection{Imaginary parts of $\ln\Lambda_{i\rmF}$}

In order to use the prescription (\ref{Imln}), the logarithmic
functions must be written in an appropriate form. Relevant forms for
$\Lambda_1$, $\Lambda_2$ follow by writing (\ref{Lambdas}) in terms
of the limiting values (\ref{limits}). This gives
\bea
\Lambda_1&=&
{(\omega-\varepsilon+\varepsilon'_{\rm max})
(\omega+\varepsilon-\varepsilon'_{\rm max})
(\omega-\varepsilon-\varepsilon'_{\rm max})
(\omega+\varepsilon+\varepsilon'_{\rm max})
\over
(\omega-\varepsilon+\varepsilon'_{\rm min})
(\omega+\varepsilon-\varepsilon'_{\rm min})
(\omega-\varepsilon-\varepsilon'_{\rm min})
(\omega+\varepsilon+\varepsilon'_{\rm min})},
\label{Lambda1a}
\\
\msp
\Lambda_2&=&
{(\omega-\varepsilon+\varepsilon'_{\rm max})
(\omega-\varepsilon-\varepsilon'_{\rm max})
(\omega+\varepsilon-\varepsilon'_{\rm min})
(\omega+\varepsilon+\varepsilon'_{\rm min})
\over
(\omega-\varepsilon+\varepsilon'_{\rm min})
(\omega-\varepsilon-\varepsilon'_{\rm min})
(\omega+\varepsilon-\varepsilon'_{\rm max})
(\omega+\varepsilon+\varepsilon'_{\rm max})}.
\qquad\qquad
\label{Lambda2a}
\eea
Although $\Lambda_3$ cannot be rewritten in terms of the factors that
appear in (\ref{Lambda1a}) and (\ref{Lambda2a}), it can be written in
a form similar to (\ref{Lambda2a}), with $\omega$ replaced by
$2\ve_k$:
\be
\fl
\Lambda_3=
{(2\ve_k-\varepsilon+\varepsilon'_{\rm max})
(2\ve_k-\varepsilon-\varepsilon'_{\rm max})
(2\ve_k+\varepsilon-\varepsilon'_{\rm min})
(2\ve_k+\varepsilon+\varepsilon'_{\rm min})
\over
(2\ve_k-\varepsilon+\varepsilon'_{\rm min})
(2\ve_k-\varepsilon-\varepsilon'_{\rm min})
(2\ve_k+\varepsilon-\varepsilon'_{\rm max})
(2\ve_k+\varepsilon+\varepsilon'_{\rm max})}.
\label{Lambda3a}
\ee
The logarithms of $\Lambda_1$, $\Lambda_2$ may be written as a sum of
terms of the form (\ref{Imln}) by identifying $(\omega-\omega_{\rm
max})/(\omega-\omega_{\rm min})$ with
$(\omega\pm\varepsilon-\varepsilon'_{\rm
max})/(\omega\pm\varepsilon-\varepsilon'_{\rm min})$ or with
$(\omega\pm\varepsilon+\varepsilon'_{\rm
min})/(\omega\pm\varepsilon+\varepsilon'_{\rm max})$. Although
$\ln\Lambda_3$ cannot be rewritten in terms of these factors, in the
neighborhood of the zeros of any of the factors in (\ref{Lambda3a}),
the vanishing factor does become of this form. As only the sign of
the imaginary part on crossing the zero is required, this suffices to
determine the sign. For example, consider the factor
$(2\ve_k-\varepsilon-\varepsilon'_{\rm max})$: this factor is zero at
the boundary of the LD region with $\varepsilon=\ve_k-\half\omega$,
$\varepsilon'_{\rm max}=\ve_k+\half\omega$, and in the neighborhood
of this boundary the factor may be approximated by $(\omega-\varepsilon-\varepsilon'_{\rm
max})$, and treated in the same manner as the corresponding factor in
$\Lambda_1$ or $\Lambda_2$.

\begin{figure}
\caption{Regions of $\omega$--$\bkm$ space ($\omega>0$) are separated
by curves corresponding to the boundaries of the regions where LD is
allowed, $\omega<\bkm$, and PC is allowed,
$\omega>(4m^2+\bkm^2)^{1/2}$. These are further separated into
regions (a)--(h) defined for a completely degenerate electron gas
with $\pf/m=1.5$. For the completely degenerate gas, LD is allowed
only in regions (b) and (c), and there is no dissipation in (a) and
(d), the electron gas completely suppresses PC in (e) and partly
suppresses PC in (f); PC has its vacuum value in (g) and (h). From
lower right to upper left the curves are:
$\omega=(\vef^2-2\pf\bkm+\bkm^2)^{1/2}-\vef$ (dashed, solid diamonds),
$\omega=\vef-(\vef^2-2\pf\bkm+\bkm^2)^{1/2}$ (dashed, solid squares),
$\omega=(\vef^2+2\pf\bkm+\bkm^2)^{1/2}-\vef$ (dot-dashed, open squares),
$\omega=\bkm$ (solid, solid circles),
$\omega=(4m^2+\bkm^2)^{1/2}$ (dotted, open circles),
$\omega=(\vef^2-2\pf\bkm+\bkm^2)^{1/2}+\vef$ (double-dot-dashed, open
triangles),
$\omega=(\vef^2+2\pf\bkm+\bkm^2)^{1/2}+\vef$ (dashed, solid triangles).
}
\label{fig:thresholds}
\end{figure}

The boundaries of the allowed regions for LD and PC are illustrated
in figure~\ref{fig:thresholds}. For $\bkm>2\pf$, one has
$\vef<\ve'_{\rm F min}$; then the upper and lower frequency
boundaries are $\omega=\ve'_{\rm F max,min}-\vef$ for LD, and
$\omega=\ve'_{\rm F max,min}+\vef$ for PC. In this case only
$\ln[(\omega\pm\vef-\ve'_{\rm F max})/(\omega\pm\vef-\ve'_{\rm F
min})]$ contribute to LD and PC, respectively, with these factors
giving an imaginary part of $\rmi\pi$ in the region where the
argument of the logarithm is negative, and zero otherwise. For
$\bkm<2\pf$ one has $\vef>\ve'_{\rm F min}$. In this case, the zero
of $\omega-\vef+\ve'_{\rm F min}$ occurs within the LD region,
separating regions (b) and (c) in figure~\ref{fig:thresholds}, and
the zero of $\omega-\vef-\ve'_{\rm F min}$ occurs within the PC
region, separating regions (e) and (f) in
figure~\ref{fig:thresholds}. It is then straightforward to determine
the signs of the imaginary parts in the various regions, and these
are listed in table~1.

\begin{table}     
\caption{Imaginary parts of RQPDFs for a completely degenerate electron gas.
}
   \begin{center}
\label{table1}
{\begin{tabular}{cccrrr}
  \hline
&&upper boundaries&
$\ln\Lambda_1$&$\ln\Lambda_2$&$\ln\Lambda_3$\\
  \hline
(c)&LD&$\omega<\vef-\ve'_{\rm F min}$,\quad$\bkm<2\pf$&$0\,$&$-i2\pi$&$0\,$\\
(b)&LD&$|\vef-\ve'_{\rm F min}|<\omega<\ve'_{\rm F
max}-\vef$&$\rmi\pi$&$-\rmi\pi$&$\rmi\pi$\\
(e)&PC&$(4m^2+\bkm^2)^{1/2}<\omega<\vef+\ve'_{\rm F
min}$,\quad$\bkm<2\pf$&$0\,$&$0\,$&$-2\rmi\pi$\\
(f)&PC&$\vef+\ve'_{\rm F min}<\omega<\vef+\ve'_{\rm F
max}$&$\rmi\pi$&$\rmi\pi$&$-\rmi\pi$
\end{tabular}}
   \end{center}
\end{table}

The imaginary parts of ${\cal P}^L(k)$ and ${\cal P}^T(k)$ may be
written down by inspection using (\ref{JL}) and (\ref{JT}),
respectively, and noting the imaginary parts in table~\ref{table1}.
These imaginary parts may also be derived from (\ref{ImPLLD}) and
(\ref{ImPTLD}) by setting the occupation number equal to unity for
$\ve<\vef$ and zero for $\ve>\vef$, and performing the integrals,
which are then elementary. The imaginary part was written down for
the longitudinal response by Jancovici (1962) and Kowalenko {\it et
al}  (1985) and these results are reproduced using (\ref{JL}) and
table~\ref{table1}.

\subsection{Alternative forms for $\Lambda_i$}

For completeness we note that the logarithmic functions may be
written in terms of the $\pm$ solutions (\ref{tpm}) and (\ref{n4});
\bea
\Lambda_1&=&{(t+t_+)(t+t_-)(t-1/t_+)(t-1/t_-)
\over(t-t_+)(t-t_-)(t+1/t_+)(t+1/t_-)}
={(|{\bi p}|+p_+)(|{\bi p}|+p_-)\over(|{\bi p}|-p_+)(|{\bi p}|-p_-)},
\nn
\\
\msp
\Lambda_2&=&{(t+t_+)(t+t_-)(t+1/t_+)(t+1/t_-)
\over(t-t_+)(t-t_-)(t-1/t_+)(t-1/t_-)}
={(|{\bi v}|+v_+)(|{\bi v}|+v_-)\over(|{\bi v}|-v_+)(|{\bi v}|-v_-)},
\nn
\\
\msp \Lambda_3&=&{(t+t_+)(t-t_-)(t+1/t_+)(t-1/t_-)
\over(t-t_+)(t+t_-)(t-1/t_+)(t+1/t_-)} ={(|{\bi v}|+v_+)(|{\bi
v}|-v_-)\over(|{\bi v}|-v_+)(|{\bi v}|+v_-)}. \qquad\qquad
\label{Lambdai} \eea The boundaries in figure~\ref{table1} are
identified as follows: the upper boundary to the PC region
corresponds to $t_\rmF=-t_-$ ($p_\rmF=-p_-$, $v_\rmF=-v_-$), the
lower boundary of region (f) corresponds to $t_\rmF=t_-$ ($\pf=p_-$,
$v_\rmF=v_-$) for $\bkm>2\pf$ and to $t_\rmF=t_+$ ($\pf=p_+$,
$v_\rmF=v_+$) for $\bkm<2\pf$; the upper boundary of region (b)
corresponds to $t_\rmF=1/t_+$ ($\pf=-p_+$, $v_\rmF=v_+$), the lower
boundary corresponds to $t_\rmF=-1/t_+$ ($\pf=p_+$, $v_\rmF=-v_+$)
for $\bkm>2\pf$ and to $t_\rmF=t_-$ ($\pf=p_-$, $v_\rmF=v_-$) for
$\bkm<2\pf$. Although useful for some other purposes, these forms
are not convenient for determining the imaginary parts because the
frequency-dependence is implicit rather than explicit, and the
prescription (\ref{Imln}) cannot be used directly.

\section{Nondegenerate thermal distribution}

The nondegenerate limit of the Fermi-Dirac distribution (\ref{FD})
applies when $\mu/T$ is large and negative, and then it becomes the
J\"uttner distribution,
\be
{\tilde n}(\varepsilon)=A\rme^{-\varepsilon/T},
\qquad
A=2\cosh(\mu/T)=\frac{\pi^2{\tilde n}_e}{2m^2TK_2(m/T)},
\label{nondegenerate}
\ee
where $K_2$ is a modified Bessel function, and ${\tilde n}_e$ is the
number density of electrons plus positrons. (The normalization
coefficient, $A$, is evaluated by setting the integral of ${\tilde
n}(\varepsilon)$ over $2\rmd^3{\bi p}/(2\pi)^3$ equal to ${\tilde
n}_e$, where the factor 2 arises from the sum over the two spin
states.)

According to Melrose and Hayes (1984), in this case the three plasma
dispersion functions $S^{(n)}(k)$ can be evaluated in terms of the
RPDF introduced by Godfrey \etal (1975), which they wrote in the form
\be
T(z,\rho)=\int_{-1}^1{\rmd v\over v-z}\,
\exp(-\rho\gamma),
\label{Tvrho}
\ee
where $z=\omega/|{\bi k}|$ is the phase speed,
$\gamma=(1-v^2)^{-1/2}$, and $\rho=m/T$ is an inverse temperature in
units of the rest energy of the electron ($\rho=1$ corresponds to
$T=0.5\times10^{10}\rm\,K$). Dissipation is described by the
imaginary part of this RPDF:
\be
{\rm Im}\, T(z,\rho)=\cases{\pi \rme^{-\rho\gamma_0}& for $|z|<1$,
\cr
\msp
0& for $|z|>1$,
\cr}
\label{ImT}
\ee
with $\gamma_0=(1-z^2)^{-1/2}$. The RQPDFs become
\bea
&&
S^{(0)}(k)={A\over\rho}\sum_\pm{\sigma_\pm\over\gamma_\pm v_\pm}
\left(
{1-v_\pm^2\over\rho}\,T'(v_\pm,\rho)
+2K_1(\rho)
\right),
\nonumber
\\
\msp
&&
S^{(1)}(k)={A\over\rho}\sum_\pm
\left[
-{T(v_\pm,\rho)\over\rho}
+{1\over v_\pm}
\left(
{1-v_\pm^2\over\rho}\,T'(v_\pm,\rho)
+2K_1(\rho)
\right)
\right],
\nonumber
\\
\msp
&&
S^{(2)}(k)=
{A\over\rho}\sum_\pm{\sigma_\pm\over\gamma_\pm v_\pm}
\bigg\{
\left(
{2\over\rho^2}+\gamma_\pm^2\right)
\left(
{1-v_\pm^2\over\rho}\,T'(v_\pm,\rho)
+2K_1(\rho)
\right)
\nonumber
\\
&&
\qquad\qquad\quad
-2\gamma_\pm^2v_\pm^2K_1(\rho)
-{2\over\rho}\,\gamma_\pm^2v_\pm
\big[
T(v_\pm,\rho)
+2v_\pm K_0(\rho)
\big]
\bigg\},
\qquad\qquad
\label{SnT}
\eea
with $T'(v_\pm,\rho)=\partial T(v_\pm,\rho)/\partial v_\pm$,
$\gamma_\pm=(1-v_\pm^2)^{-1/2}$, and with $A$ given by
(\ref{nondegenerate}). The sign $\sigma_\pm=\ve_\pm/|\ve_\pm|$ is
needed in the LD region to take account of the fact that $\ve_\pm$
can be negative while $\gamma_\pm=|\ve_\pm|/m$ are positive by
definition. The  sign $\sigma_\pm$ is replaced by unity in the DL and
PC regions. The RQPDFs (\ref{SnT}) characterize the response of a
nondegenerate thermal electron gas when RQ effects are included.

The interpretation of $v_\pm$ and $\gamma_\pm$ is different in the LD
and PC regimes. In the LD region one has
\be
v_\pm=\frac{\omega}{\bkm}\,
\frac{\ve_k\pm\bkm^2/2\omega}{\ve_k\pm\omega/2},
\qquad
\gamma_\pm=(\ve_k\pm\half\omega)/m,
\label{vpm}
\ee
with $\ve_k$ defined by (\ref{vek}). In the nonquantum limit, one has
$\ve_k\gg\omega/2,\bkm^2/2\omega$, implying $v_\pm\to\omega/\bkm=z$.
Then (\ref{SnT}) reproduces the known non-quantum limit (Melrose and
Hayes 1984):
\bea
&&
S^{(0)}(k)={\omega A\over mz}
\big[zT(z,\rho)+2K_0(\rho)
\big],
\nonumber
\\
\msp
&&
S^{(1)}(k)={2A\over\rho^2z}
\big[
-zT(z,\rho)
+(1-z^2)\,T'(z,\rho)
+2\rho K_1(\rho)
\big],
\nonumber
\\
\msp
&&
S^{(2)}(k)={\omega A\over mz}
\big[\gamma_0^2zT(z,\rho)+2\gamma_0^2(1+z^2)K_0(\rho)+K_2(\rho)
\big].
\qquad\qquad
\label{SnNQ}
\eea
In comparing (\ref{SnT}) with (\ref{SnNQ}), it is apparent that the
phase speed, $z=\omega/|{\bi k}|$, in the nonquantum case is replaced
by two functions $v_\pm$ that include the effect of the quantum
recoil, which has opposite signs for emission and absorption. Thus,
in the LD region, $v_\pm$ are interpreted as resonant phase speeds
for stimulated emission and true absorption, which differ due to the
quantum recoil, and $m\gamma_\pm$ are interpreted as the energies of
the electron before and after emission of a wave quantum,
respectively. In the PC regime, $m\gamma_\pm=\ve_k\pm\half\omega$ are
interpreted as the energies of the created electron and positron.

In the DL regime the RQPDFs must be real. In this case, the $v_\pm$
are complex conjugates of each other. With $\sigma_\pm=1$ and
$T(v^*,\rho)=T^*(v,\rho)$, the sum over $\pm$ in (\ref{SnT}) leads to
a real expression, as required.

Comparison of the imaginary parts for the response functions obtained
from the imaginary parts of the RQPDFs with those obtained by
imposing the causal condition directly provides a check on both
results. Here the imaginary parts that are to be compared are those
obtained by inserting the imaginary parts of the $S^{(n)}(k)$ into
the expressions (\ref{calPL}) and (\ref{calPT}) for the longitudinal
and transverse response, and those obtained by evaluating the
integrals in (\ref{ImPLLD}), (\ref{ImPTLD}) and (\ref{ImPLPC}),
(\ref{ImPTPC}) for the nondegenerate distribution
(\ref{nondegenerate}). The results agree. In particular, Tystovich
(1961)
wrote down explict expressions for dissipation due to LD and PC in a
nondegenerate electron gas, which our derivation reproduces: the
spurious factor of two is present only in Tsytovich's calculation of
the vacuum contribution.

\section{Applications}

RQ effects become important for dissipation and dispersion in plasmas
only under extreme conditions, such as the early Universe,
quark-gluon plasmas and the interiors of compact stars. The
degeneracy condition (temperature less than chemical potential) is
relevant to only the last of these, and we concentrate on this case,
emphasizing the role of PC. First we comment on LD.

LD in the non-quantum limit is possible only for subluminal waves,
$\omega<\bkm$, and this is also the case when RQ effects are
included. This precludes LD of transverse waves, which are
superluminal. Dissipation and dispersion associated with LD may be
treated nonrelativistic provided not only that the particles are
nonrelativistic, but also that the waves are subluminal. The RQ
recoil term changes the classical resonance condition to $\omega-{\bf
k}\cdot{\bf v}\pm(\omega^2-\bkm^2)/2m\gamma$, for emission and
absorption, whereas in nonrelativistic theory, the recoil term is
$\mp\bkm^2/2m$. The difference is unimportant for nonrelativistic
particles and subluminal waves, but is important for superluminal
waves and for waves with near vacuum dispersion, $\omega\approx\bkm$.
An implication is that the widely used response functions of Lindhard
(1954), which were derived using nonrelativistic quantum mechanics,
may lead to unreliable results for waves with $\omega\gapprox\bkm$.
We are currently investigating this point.

The contribution of the plasma process to neutrino emission from the
cores of compact stars depends on the dispersive properties of the
degenerate gas, and Braaten (1991) pointed out that earlier authors,
following Baudet, Petrosian and Salpeter (1967), had used inaccurate
forms for the dispersion relations. A controversial point was raised
by Braaten (1991), who criticized the claim by Baudet, Petrosian and
Salpeter (1967), and subsequent authors, that PC needs to be taken
into account in sufficiently hot and dense plasmas: if PC is allowed
then photons decay into pairs much faster than they would decay into
neutrinos. This point was discussed further by  Itoh \etal (1992)
and Braaten and Segel (1993), who also concluded that PC is
forbidden in a completely degenerate electron gas. In more recent
discussions (Ratkovi$\acute c$, Dutta and Prakash 2003; Dutta,
Ratkovi$\acute c$ and Prakash 2004; Koers and Wijers 2005; Jaikumar,
Gale and Page 2005), the approximations of Braaten and Segel (1993)
to the dispersion functions have been used.  Our results show that
PC is allowed in a completely degenerate electron gas, and the
reason that our results differ from those of these earlier authors
can be understood as follows.

Itoh \etal (1992) considered wave quanta at the cutoff frequency,
$\omega_c$, and argued that although one can have $\omega_c>2m$ in a
superdense plasma, the actual threshold for PC is $2\vef$, and one
cannot have $\omega_c>2\vef$. The higher threshold is because all
electron states below the Fermi energy are occupied, and for
$\bkm=0$ the electron and positron energies are equal. This argument
is consistent with our results, but it applies only at $\bkm=0$. PC
is forbidden in region (e) in figure~\ref{fig:thresholds}, and this
region shrinks as $\bkm$ increases. A dispersion curve for
transverse waves that starts at $\omega_c>2m$ for $\bkm=0$ is
necessarily in region (e) for sufficiently small $\bkm$, but then
necessarily enters region (f), where PC is allowed, before
approaching the light line asymptotically. Braaten and Segel (1993)
made approximations to the wave dispersion by neglecting the quantum
recoil: specifically, if one combines the denominators in
(\ref{calP0}), the common factor may be written as
$(ku)^2-(k^2/2m)^2$, and Braaten and Segel (1993) argued that for
practical purposes one can neglect the $(k^2/2m)^2$ term. However,
near the cutoff $\bkm\to0$, one has
$(ku)^2-(k^2/2m)^2\to\omega_c^2\gamma^2(1-\omega_c^2/4m^2\gamma^2)$,
and their approximation requires $\omega_c\ll2m\gamma$, which is not
satisfied for all $1<\gamma<\vef/m$ in a superdense plasma. This
approximation effectively excludes dispersion due to PC, and it is
inconsistent to use it to argue that PC cannot occur. Our results
show that PC does occur over a limited range of $\bkm$ in a
completely degenerate electron gas, but not in region (e) in
figure~\ref{fig:thresholds}, due to exact cancellation of the vacuum
contribution to PC by the electron gas. For a partially degenerate
electron gas, even in region (e) the cancellation is not exact, and
PC occurs.

The original argument of Braaten (1991) against PC was based on mass
renormalization of the electron, arguing that this suppresses the
cutoff frequency, keeping it below the PC threshold. We do not
comment specifically on this argument here. Our conclusion is that
the arguments by Itoh \etal (1992) and Braaten and Segel (1993) that
neglected mass renormalization do not negate the original argument of
Baudet, Petrosian and Salpeter (1967) that PC needs to be taken into
account when considering the plasma process for neutrino emission in
a superdense plasma.

\section{Conclusions}

In this paper we discuss the properties of RQPDFs for an isotropic,
unmagnetized plasma. The dispersion is related to the dissipation,
which includes the familiar Landau damping (LD), modified by the
quantum recoil, and one-photon pair creation (PC). It is necessary to
treat the dissipation in the LD and PC regimes differently, and to
interpret them differently. LD has the same interpretation as in a
non-quantum plasma, except that the resonance at the phase speed,
$z=\omega/\bkm$, is replaced by resonances at two speeds, $v_\pm$,
and corresponding energies, $\ve_\pm$, given by (\ref{vpm}) and
interpreted as the resonant values for induced emission and true
absorption when the quantum recoil is included. Dissipation due to PC
in the electron gas has the opposite sign to LD and a different
interpretation: PC exists in the vacuum, due to the imaginary part of
the vacuum polarization tensor (\ref{Imvac}), and the presence of an
electron gas partly suppresses PC due to the Pauli exclusion
principle.

An objective in this paper is to relate dissipation and dispersion by
deriving the dissipation from the imaginary parts of the RQPDFs,
which requires that the imaginary parts be determined explicitly. In
particular, the logarithmic RQPDFs that appear for a completely
degenerate electron gas acquire an imaginary part of $\pm \rmi\pi$
when their arguments becomes negative, and a prescription is needed
to determine the sign of this imaginary part uniquely. We start from
the DL region, where the imaginary part is necessarily zero, and
analytically continue into the regions where LD and PC are allowed.
We show that the Landau prescription leads to a relatively simple
prescription for identifying the sign of the imaginary part acquired
when the argument of the logarithm changes sign. We compare our
results with existing expressions for the imaginary parts derived in
other ways, and find agreement provided that some minor errors are
corrected. Specific errors identified are a spurious multiplicative
factor in Jancovici's (1962) transverse response function and a
factor of two in the expression derived by Tsytovich (1961) for the
vacuum contribution to dissipation due to PC.

In the absence of any plasma, dissipation due to PC is determined by
the imaginary part of the vacuum polarization tensor for
$\omega>(m^2+\bkm^2)^{1/2}$, and is zero otherwise. The presence of
an electron gas tends to suppress PC, and the presence of a
completely degenerate electron gas can completely suppress PC.
Complete suppression at a given $\omega,\bkm$ occurs if all potential
states for the created electron are below the Fermi level. Although
it was argued by Itoh \etal (1992) and Braaten and Segel (1993) that
PC cannot occur in a superdense plasma, where the cutoff frequency
exceeds the PC threshold $2m$, we show this is not the case for at
least a range of $\bkm\ne0$. The earlier arguments of Baudet,
Petrosian and Salpeter (1967) on the implications of PC remain valid
and need to be taken into account in detailed analyses (e.g.,
Jaikumar, Gale and Page 2005).

\section*{\it Acknowledgments}
\noindent We thank Qinghuan Luo for helpful comments on the manuscript.

\section*{References}
\begin{harvard}

\item[]
{Baudet, G, Petrosian, V, and Salpeter E E}
{1971}
{\it Astrophys. J.} {\bf 150}, 979

\item[]
{Berestetskii, V B, Lifshitz, E M and Pitaevskii, L P}
{1971}
{\it Relativistic Quantum Theory}, Pergamon
Press

\item[]
Braaten, E
1991
{\it Phys. Rev. Lett.} {\bf 66} 1655

\item[]
Braaten, E
1992
plasmino annihilation,
{\it Astrophys. J.} {\bf 392} 70

\item[]
Braaten, E, and Segel, D
1993
{\it Phys.\ Rev.\ D} {\bf 48} 1478

\item[]
Delsante, A E and Frankel, N E
1980
{\it Ann.\ Phys.\ (NY)} {\bf 125} 135

\item[]
{Dutta, S I, Ratkovi$\acute c$, S, and Prakash, M}
{2004}
{{\it Phys.\ Rev.\ D\/} {\bf 69} 023005}

\item[]
{Godfrey, B B , Newberger, B S and Taggart, K A}
{1975}
{{\it IEEE Trans. Plasma Sci.} {\bf PS-3} 60}

\item[]
Hakim, R
1978
{\it Riv.\ Nuovo Cim}. {\bf 1} 1

\item[]
Hakim, R and Heyvaerts, J
1978
{\it Phys. Rev. A\/} {\bf 18} 1250

\item[]
Hakim, R and Heyvaerts, J
1980
{\it J. Phys. A\/} {\bf 13} 2001

\item[]
{Hayes, L M, and Melrose, D B}
{1984}
{{\it Aust.\ J Phys}. {\bf 37} 615}

\item[]
Itoh, N, Mutoh, H, Hikita, A, and Kohyama, Y
1992
process for strongly degenerate electrons
{\it Astrophys. J}. {\bf 395} 622

\item[]
{Jaikumar, P, Gale, C, and Page, D}
{2005}
{{\it Phys.\ Rev.\ D\/} {\bf 72} 123004}

\item[]
{Jancovici, B}
{1962}
{{\it Nuovo Cim}. {\bf 25} 428}

\item[]
{Koers, H B J, and Wijers, R A M J}
{2005}
{{\it Mon.\ Not.\ Roy.\ Astron.\ Soc.} {\bf 364} 934}

\item[]
{Kowalenko, V, Frankel, N E and Hines, K C}
{1985}
{{\it Phys.\ Rep.} {\bf 126} 109}

\item[]
{Lindhard, D J}
{1954}
{{\it Mat.\ Fys.\ Medd.\ Dan.\ Vid.\ Selsk}.\ {\bf 28} no.\ 8, p 1}

\item[]
{Melrose, D B and Hayes, L M}
{1984}
{{\it Aust. J. Phys}. {\bf 37} 639}

\item[]
{Ratkovi$\acute c$, S, Dutta, S I, and Prakash, M}
{2003}
{{\it Phys.\ Rev.\ D\/} {\bf 67} 123002}

\item[]
{Sivak, H D}
{1985}
{Ann.\ Phys. {\bf 159} 351}

\item[]
{Tsytovich, V N}
{1961}
{\it Sov.\ Phys.\ JETP} {\bf 13} 1249

\item[]
{Williams, D R M  and Melrose, D B}
{1989}
{{\it Aust.\ J.\ Phys}. {\bf42} 59}

\end{harvard}

\end{document}